\documentclass[twocolumn,preprintnumbers,pra]{revtex4}
\pdfoutput=1
\usepackage{graphicx}
\usepackage{times}

\usepackage[hypertex]{hyperref}
\usepackage{slashed}

\begin{document}

\title{Quark mass uncertainties revive KSVZ axion dark matter}
\author{Matthew R. Buckley and Hitoshi Murayama}
\affiliation{Department of Physics, University of California,
  Berkeley, CA 94720, USA} \affiliation{Theoretical Physics Group,
  Lawrence Berkeley National Laboratory, Berkeley, CA 94720, USA}
\date{\today}

\begin{abstract}
  The Kaplan-Manohar ambiguity in light quark masses allows for a
  larger uncertainty in the ratio of up to down quark masses than naive
  estimates from the chiral Lagrangian would indicate. We show that it allows for a
  relaxation of experimental bounds on the QCD axion, specifically KSVZ axions in the $2-3~\mu$eV mass range composing 100\% of the galactic dark matter halo can evade the experimental limits placed by the ADMX collaboration.
\end{abstract}
\maketitle

\section{Introduction \label{sec:intro}}

In the energy budget of the universe, it appears that approximately
one quarter of the cosmos is composed of unidentified weakly
interacting particles. The as-yet-unseen axion, originally proposed to
solve the strong CP problem, has couplings to normal matter suppressed
by a large energy scale.  As such it provides an ideal dark matter
candidate. There are two possible ranges of axion masses that would
result in sufficient axionic matter density in the early universe to
provide most, if not all, of the currently seen $\Omega_{DM}$. A mass
scale of $\sim \mu$eV would provide cold dark matter through coherent
oscillation of the axion field, while an eV mass would produce axions
as thermal relics \cite{Kamionkowski:1997zb}.

From measurements of the neutron dipole moment, it is known that the
strong interaction preserves CP symmetry. This presents a conundrum
for quantum field theory, as no gauge or global symmetry prohibits the
CP violating $\theta$-term in the Lagrangian:
\begin{equation}
{\cal L}_\theta = -\frac{g_s^2}{64\pi^2}\theta\epsilon_{\mu\nu\rho\sigma}G^{a\mu\nu}G^{a\rho\sigma}=-\frac{g_s^2}{32\pi^2} \theta G^a\tilde{G}^a. \label{eq:theta}
\end{equation}
This term is a total derivative, and therefore one is
tempted to ignore it. However the topological structure of $SU(3)$ allows
Eq.~(\ref{eq:theta}) to contribute through instanton effects. Indeed,
the experimental fact that there is not a meson with mass $<
\sqrt{3}m_\pi$ which can be identified as the Nambu-Goldstone (NG)
boson of a spontaneously broken $U(1)_A$ global symmetry of the strong
interaction implies that there is no axial symmetry, and therefore we
must take the presence of the $\theta$ term seriously
\cite{Weinberg:1975ui}.

Experiments limit $\theta \lesssim 10^{-10}$
\cite{Altarev:1981zp}\cite{Pendlebury1984}\cite{Baker:2006ts}, in
contrast to naive expectations that $\theta$ should be ${\cal O}(1)$. The
mystery deepens when one considers that the physical observable is not
$\theta$ itself, but $\bar{\theta}=\theta-\arg\det M$ where $M$ is the
quark mass matrix. It is difficult to see how $\theta$, originating
with the strong interaction, could cancel to such high precision with
the phases of quark masses, originating in the electroweak sector.

A favored approach to eliminate Eq.~(\ref{eq:theta}) is to seek a
dynamical method to set $\theta$ to zero, by introducing a new degree of freedom: the axion
\cite{Peccei:1977ur}\cite{Peccei:1977hh}\cite{Weinberg:1977ma}\cite{Wilczek:1977pj}. In
brief, an axion model requires a new global Peccei-Quinn (PQ)
symmetry, at least one scalar field, and strongly interacting fermions
with PQ charges. The scalar field acquires a vev, breaking the PQ
symmetry, and the resulting Nambu-Goldstone boson becomes the
axion. Through triangle anomalies, the axion mixes with mesons
(resulting in a non-zero mass), and thus couples to photons, nucleons,
and leptons.

The axion mass and couplings are suppressed by powers of the axion decay constant $f_a$. From astrophysical constraints $f_a \gtrsim4 \times 10^8$~GeV \cite{Raffelt:2006cw}.
For $f_a \sim 10^{12}$~GeV, the axion could provide cold dark matter with $\Omega \sim {\cal O}(1)$ with very weak coupling to the Standard Model particles. Many searches for axions have been performed, so far
all have returned negative results \footnote{The possible discovery of
  a pseudoscalar particle by PLVAS \cite{Zavattini:2005tm} from the
  phase rotation of light \cite{Maiani:1986md}\cite{Raffelt:1988}
  cannot be the QCD axion, as the relationship between mass and
  coupling to photons falls outside the allowed range for axions (see
  Eq.~(\ref{eq:Gformula})).}.  
  
We focus here on recent bounds placed
on the axion to photon coupling by the ADMX experiment \cite{Asztalos:2001tf}, 
and the derived limits on the axion decay constant $f_a$. The ADMX 
experiment is of particular interest as it has placed the most restrictive bounds
on the axion-photon coupling in the regime $f_a \sim
10^{12}$~GeV. Assuming that the dark matter in our galactic halo is
comprised completely of axions and has a density $\rho_{DM} = 0.45$~GeV/cm$^3$, ADMX excludes the KSVZ axion model with $E/N=0$ \footnote{The KSVZ model was originally put forth by Kim
\cite{Kim:1979if} and separately by Shifman, Vainshtein, and Zakharov
\cite{Shifman:1979if}. The parameter $E/N$ is introduced in section \ref{sec:formalism}.} in the axion mass range $1.9-3.3~\mu$eV. However, it was noted
\cite{Chang:1993gm}\cite{Moroi:1998qs} that uncertainty in the masses
of the light quarks due to the Kaplan-Manohar ambiguity
\cite{Kaplan:1986ru} introduces larger than predicted uncertainties in
the axion-photon coupling. We shall show that this allows
the KSVZ model to evade the experimental bounds from ADMX.

In section \ref{sec:formalism}, the full effects on axion mass and
coupling formulae from the Kaplan-Manohar ambiguity are calculated for
general axion models. Section \ref{sec:exp} then analyzes what regions
of parameter space may be excluded by experiments in light of these
ambiguities. We conclude in section \ref{sec:conc}. Details of the
calculations as well as explicit formulae are provided in appendix A, while the analysis of two quark flavors is shown in appendix B.

\section{Axion Theory \label{sec:formalism}}

We start with the most general possible axion model \footnote{We
  ignore the possibility of a string theory axion; though a similar
  analysis is possible in such models, the details of the calculations
  would differ. In addition, it appears difficult to reconcile the the cosmological constraints with the
  scale of the axion decay constant $f_a\sim M_{Pl}$ preferred by string theory. See Section \ref{sec:exp} and reference \cite{Svrcek:2006yi} for a
  discussion of this.}; containing at least one complex scalar field
$\sigma$ and with chiral fermions $\psi$ charged under a new global PQ
symmetry. At least some of the $\psi$ must have non-trivial $SU(3)$
quantum numbers. 

For a specific example, we consider the KSVZ model \cite{Kim:1979if}\cite{Shifman:1979if}, in which the
scalar field $\sigma$ has PQ charge $+2$, and the fermions are new
$SU(3)$ triplets. $\psi_L$ has PQ charge $+1$, and $\psi_R$ has charge
$-1$. The assignment of PQ charges precludes explicit mass
terms for the fermions, but allows Yukawa interactions
$\bar{\psi}_L\sigma\psi_R$. In this model, the SM fields are not charged under PQ
symmetry. Another popular model is the DFSZ axion, proposed by Dine,
Fischler, and Srednicki \cite{Dine:1981rt} and separately by
Zhitnitsky \cite{Zhitnitsky:1980tq}. Here two Higgs doublets
provide mass to the Standard Model particles, with another new
scalar field to break the PQ symmetry. The SM fermions, as well as the
three scalars, are all charged under a global PQ symmetry. As the SM
fermions naturally fall into complete $SU(5)$ multiplets, this is a
simple example of an axion model with particle content consistent with grand unification (GUT).

We shall perform explicit calculations in the KSVZ model, and note that the low energy axion-photon phenomenonology for DSFZ is identical with the parameter $E/N$ (which we introduce shortly) set to $8/3$. In the KSVZ model, the high energy Lagrangian for fields
with PQ charge is
\begin{equation}
{\cal L}_{\mbox{ \small Kin.}}-h(\bar{\psi}_L \sigma \psi_R +\mbox{h.c.})-V(\sigma)-\frac{g_s^2}{32\pi^2}\bar{\theta} G^a\tilde{G}^a. \label{eq:highELag}
\end{equation}
The scalar $\sigma$ acquires a vev $v/\sqrt{2}$ through some
dynamics of the scalar potential. The fermions get PQ-breaking mass
terms, and the remaining massless mode $\phi$ of the scalar $\sigma$
appears as a phase to this term:
\begin{equation}
h(\bar{\psi}_L \sigma \psi_R +\mbox{h.c.}) \to \frac{1}{\sqrt{2}}hv(\bar{\psi}_L\psi_R e^{i\phi/v}+\mbox{h.c.}) \label{eq:psimass}
\end{equation}
By judicious choice of a chiral rotation on $\psi$, we can eliminate
the $\bar{\theta}$ angle in Eq.~(\ref{eq:highELag}). To do so, we
recall from the Fujikawa measure of the path integral that a rotation
of $e^{i \alpha\gamma_5}$ on a Dirac spinor contributes
\begin{equation}
\frac{g_s^2}{8\pi^2} \alpha G^a \tilde{G}^b \mbox{Tr}(t^at^b) \label{eq:anomaly}
\end{equation}
to the Lagrangian (here the $t^a$ are the generators of the
representation of $SU(3)$ to which $\psi$ belongs). Defining
$2\sum\mbox{Tr}(t^at^b) = N\delta^{ab}$, where the sum runs over all
$\psi$ with PQ charge ($N=1$ for one Dirac $SU(3)$ triplet), and the axion decay constant $f_a \equiv v/N$, we see that a
rotation of
\begin{equation}
\psi \to \exp\left(-\frac{i\phi\gamma_5}{2Nf_a} \right)\psi \label{eq:psirotation}
\end{equation}
sends $\bar{\theta} \to \bar{\theta}+\phi/f_a$ in
Eq.~(\ref{eq:highELag}). It is important to note that, should any of
the $\psi$ have $U(1)_{EM}$ charges, the transformation
Eq.~(\ref{eq:psirotation}) will also add a $F\tilde{F}$ term to the
Lagrangian:
\begin{equation}
{\cal L} \to {\cal L} -\frac{e^2}{32\pi^2}\left(\frac{E}{N}\right)\frac{\phi}{f_a}F\tilde{F}. \label{eq:aFF1}
\end{equation}
Here $E=2\sum Q^2$, where again the sum runs over all PQ-charged Dirac fermions. Since $\phi$ is not a
constant, this term is not a total derivative, and so cannot be ignored.

As will be shown, at low energies $\phi$ will get a vev, 
$\langle \phi \rangle = -f_a\bar{\theta}$,
eliminating the constant $\bar{\theta}$ term in the Lagrangian. The
axion then is the excitation of the $\phi$ field, $a = \phi-\langle
\phi \rangle$. At energies far below $f_a$, we can integrate out the
heavy degrees of freedom, leaving the effective Lagrangian
\begin{equation}
{\cal L}_{SM}+\frac{1}{2}\partial_\mu a\partial^\mu a-\frac{g_s^2}{32\pi^2}\frac{a}{f_a}G^a\tilde{G}^a-\frac{e^2}{32\pi^2}\frac{E}{N}\frac{a}{f_a}F\tilde{F}. \label{eq:axionLag}
\end{equation}

To derive the axion coupling to photons, we eliminate the coupling of
axions to gluons through rotation of the light quark fields. Later, as
we move below the QCD scale, this will result in mixing between axions
and NG mesons of the broken chiral $SU(2)\times SU(2)$. Explicitly, we
rotate
\begin{equation}
q \to \exp\left(\frac{ia}{f_a} \frac{ \gamma_5}{2\times 3}\right)q \label{eq:qrotation}
\end{equation}
(where $q=u,d,s$). The factor of three is the number of quark flavors
being rotated, with an additional factor of two for left/right handedness.

After the rotation Eq.~(\ref{eq:qrotation}), the quark-axion sector of
the Lagrangian Eq.~(\ref{eq:axionLag}) is
\begin{eqnarray}
 \lefteqn{ {\cal L}  =  i\bar{q} \slashed{D}q+\frac{1}{2}(\partial_\mu a)^2+\frac{1}{6f_a}\bar{q}\gamma^\mu\gamma_5q\partial_\mu a} & &  \label{eq:qaLag} \\
  & & +\left(\bar{q}_L M e^{ia/3f_a}
  q_R+\mbox{h.c.}\right)-\frac{e^2}{32\pi^2}\left(\frac{E}{N}-\frac{4}{3}\right)\frac{a}{f_a}F\tilde{F} \nonumber
\end{eqnarray}
here $M=\mbox{diag}(m_u,m_d,m_s)$ is the light quark mass matrix. 

We now consider the Lagrangian below the QCD scale, where the quarks
have hadronized into mesons. In doing so the kinetic mixing terms
vanish, as we have (by design) rotated all three light quarks by the
same phase and there is no flavor singlet in the chiral Lagrangian.

To determine the axion-photon coupling and the axion mass we use the
chiral Lagrangian, which contains powers of the quark mass matrix
$M$. To include the axion in this, we take our lead from
Eq.~(\ref{eq:qaLag}) and simply rotate $M$ by the phase $e^{ia/3f_a}$
in the Lagrangian.

Let us now consider the form of the chiral Lagrangian, ignoring for
the moment the axion phase rotation. To leading order in $M$, the mass
term is simply
\begin{equation}
{\cal L}_{mass} = \frac{1}{2}\mu f_\pi^2 \mbox{Tr}(M\Sigma)+\mbox{h.c.} \label{eq:chiralmass1}
\end{equation}
where $\Sigma = \exp[2i \pi^a T^a/f_\pi]$ ($a=1,\ldots
8$) is the meson field, $T^a$ are the generators of the adjoint representation of $SU(2)$, $\mu$ is an undetermined constant, and $f_\pi = 93$~MeV. As is
well known, one can expand out the exponential $\Sigma$ field and, by
taking the terms of ${\cal O}(\pi^a)^2$, find the masses of the NG
meson octet in terms of $\mu$ and the light quark masses (Eqs.~(\ref{eq:PiEtaAmass}) and (\ref{eq:Kmass})). Comparison
with experiment then allows one to measure the ratio of masses,
$m_u/m_d \equiv z$ and $m_u/m_s\equiv w$. The resulting value for $z$
($z=0.56$) we shall refer to as the Weinberg value for $m_u/m_d$
\cite{Weinberg:1977hb}.

Adding in the axion field, the procedure is identical. One simply
expands both $\Sigma$ and $e^{ia/3f_a}$ to second order in the axion
and meson fields and reads off their masses directly. As will be
shown in more detail later in this paper, doing so results in axion
masses and couplings in terms of $\mu$, $z$, $w$, $f_a$, and
$E/N$. Using the Weinberg value, one can then place limits on the
scale $f_a$ from null results of axion searches.

It is also at this point that the $\phi$ field (of which the axion is the excitation) develops a vev. Expanding $\Sigma$ and considering the constant term, we see that the $\phi$ potential is
\begin{eqnarray}
V(\phi) & = & -\frac{\mu f_\pi^2}{2} \mbox{Tr}(M)\exp\left[ \frac{i}{3f_a}\left(\bar{\theta}+\frac{\phi}{f_a} \right)\right]+\mbox{h.c.} \nonumber \\
& = & -\mu f_\pi^2\mbox{Tr}(M)\cos\left[\frac{1}{3f_a}\left(\bar{\theta}+\frac{\phi}{f_a} \right)\right]. \label{eq:Vphi}
\end{eqnarray} 
This is minimized when $\phi = \langle \phi \rangle = -\bar{\theta}f_a$, which justifies our expansion about this field value in parametrizing the axion $a$. 

Returning to the chiral Lagrangian, there is a significant ambiguity in the value of $z$ that is
not apparent from consideration of Eq.~(\ref{eq:chiralmass1})
\cite{Kaplan:1986ru}. A naive uncertainty on the Weinberg value
of $z$ would be $0.56\pm 0.05$ \cite{Gasser:1982ap}. However, the chiral Lagrangian
contains terms of higher order in $M$. Of particular interest is the 
Kaplan-Manohar term:
\begin{equation}
{\cal L}_{mass} = \frac{\mu f_\pi^2}{2} \left[\mbox{Tr} M \Sigma+\frac{\delta}{m_s}  \mbox{Tr}[\det(M) M^{-1} \Sigma^\dag]^\dag +\mbox{h.c.} \right]. \label{eq:chiralmass2}
\end{equation}
This term approximately remaps
\begin{equation}
\mu m_u \to \mu m_u + \beta m_dm_s \label{eq:kaplanremap}
\end{equation}
($m_d$ and $m_s$ have similar transformations) where $\beta = \mu
\delta/m_s$. By naive dimensional analysis, the Kaplan-Manohar term should have a coefficient $\mu^2 f_\pi^2/\Lambda_{\chi}^2$, where $\Lambda_{\chi} = 4\pi f_\pi$ is the strong scale \cite{Georgibook}. Comparison with our choice of normalization gives
\begin{equation}
\delta \sim \frac{\mu m_s}{(4\pi)^2f_\pi^2}. \label{eq:deltamag}
\end{equation}
As $\mu m_s \approx 0.24$~GeV$^2$, $\delta$ is $\sim 0.2$.

As shown in appendix \ref{sec:2flavor}, in the limit of two flavors, the dominant effect of the Kaplan-Manohar term is the remapping of Eq.~(\ref{eq:kaplanremap}), while it's inclusion in the three flavor case has more complicated results. This remapping means that the ratio $z$ is essentially unconstrained
by considerations of the meson masses alone.

While the chiral Lagrangian alone cannot determine $z$, other methods 
are of more use, primarily lattice QCD and higher order chiral perturbation theory. The range on the ratio $z$ adopted by the Particle Data Group
is \cite{Manohar:2004fp}\footnote{It should be noted that an error in the PDG resulted in an out of date plot for the allowed range of $m_u$ vs. $m_d$ in reference \cite{Manohar:2004fp}. The correct figure can be found at http://pdg.lbl.gov/2006/figures/quark\_masses2.eps. We thank Michael Barnett for clarifying this issue.}
\begin{equation}
0.3 \leq z \leq 0.6. \label{eq:zlimits} 
\end{equation}
The calculation of the range of $z$ has considerable uncertainty associated with it, for example reference \cite{Maltman:1989sx} quotes a value up to $0.8$. To allow for this lack of a general consensus in the community, we therefore also include a more conservative estimate of the $z$ range in our calculations:
\begin{equation}
0.2 \leq z \leq 0.7. \label{eq:zlimits2} 
\end{equation}
This corresponds to a $20\%$ contribution from next-to-leading order terms in the chiral Lagrangian \cite{Kaplan:1986ru}, and is consistent with several measurements of the mass ratio \cite{Maltman:1989sx}\cite{Leutwyler:1989pn}. 

\begin{figure}[h]
\centering
\includegraphics[width=\columnwidth]{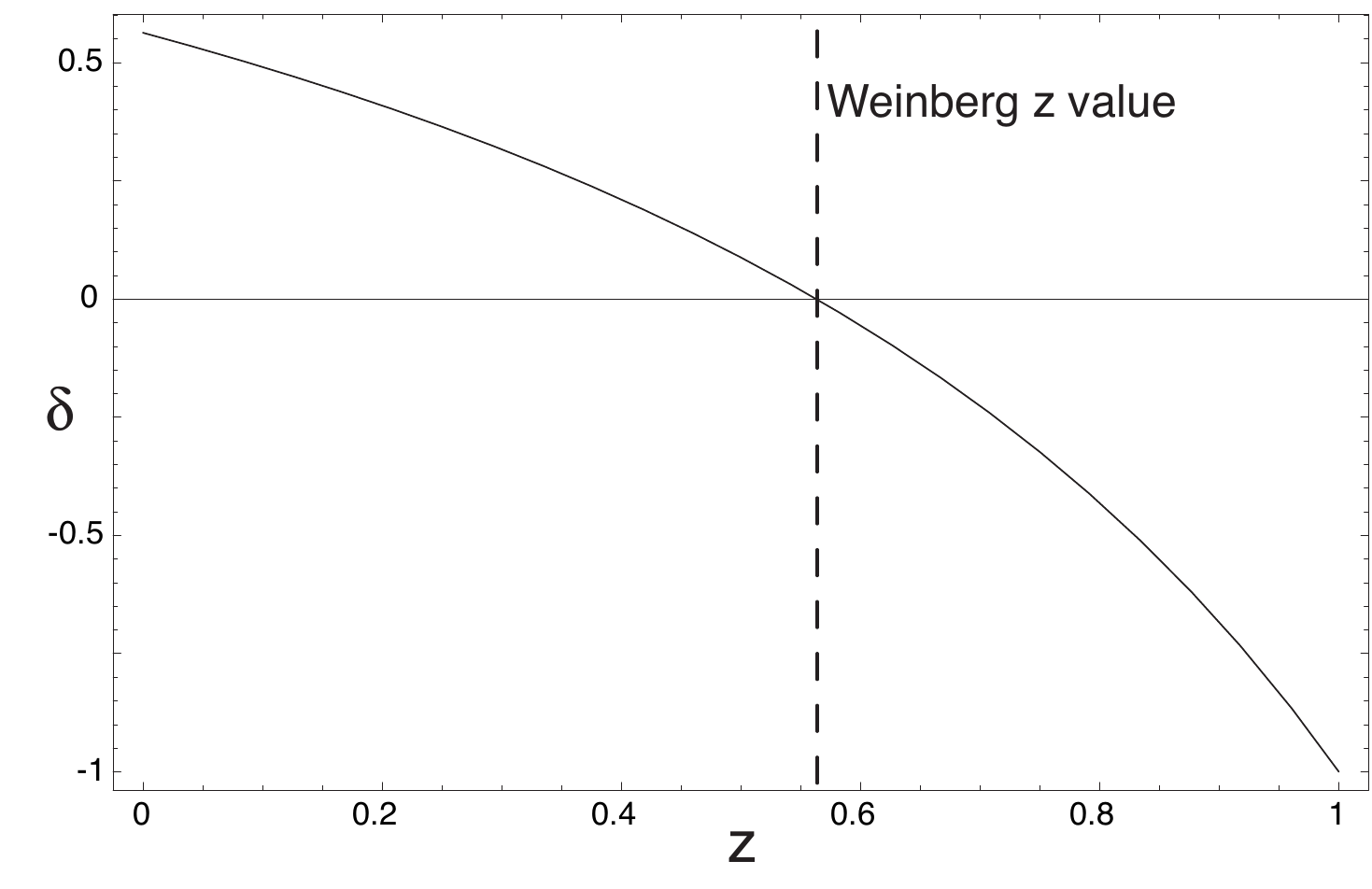}

\caption{Coefficient $\delta$ of Kaplan-Manohar term in the chiral
  Lagrangian as a function of $z$, the ratio of up to down quark
  masses. The Weinberg value for $z=0.56$ corresponds to
  $\delta=0$. The functional form of $\delta$ is given in Eq.~(\ref{eq:DeltaWMu}).} \label{fig:delta}
\end{figure}

Once $z$ is determined, the ratio $w$, parameters $\delta$ and $\mu m_u$ can be calculated as functions of $z$ and the physical meson masses (Eq.~(\ref{eq:DeltaWMu})) \cite{Kaplan:1986ru}.  As detailed in appendix A, for simplicity we expand the relevant meson masses to leading order in $w$, exact solutions differ by less than $3\%$ from this approximation over the allowed range of $z$. Additionally, comparison with reference \cite{Kaplan:1986ru} shows variation of up to 10\% in the solution for $w$ as a function of $z$. This can be attributed to additional higher order terms included in the analysis of  reference \cite{Kaplan:1986ru}. The solution of $\delta$ as a function of $z$ is shown in Fig.~\ref{fig:delta}. As expected, we find that $\delta$ is a number of less than order one, and the Weinberg value of $z$ corresponds to $\delta=0$. 

Applying the rotation
$M \to Me^{ia/3f_a}$ in Eq.~(\ref{eq:chiralmass2}) to include the axion, we expand and take
the terms quadratic in the fields. From this, one obtains the axion/neutral meson mass matrix ${\cal M}^2$ (Eq.~(\ref{eq:PiEtaAmass})).

To complete the calculation of the axion-photon coupling, we recall
that both the $\pi^0$ and $\eta$ couple to photons through triangle
anomalies, and, from Eq.~(\ref{eq:PiEtaAmass}), have mixings with the
axion. The couplings to photons for the light mesons are
\begin{eqnarray}
\lefteqn{\frac{1}{4}G_{\pi\gamma\gamma} \pi^0F\tilde{F}+ \frac{1}{4}G_{\eta\gamma\gamma} \eta F\tilde{F} =} & & \nonumber \\ & & +\frac{e^2}{32\pi^2}(2)\frac{\pi^0}{f_\pi}F \tilde{F}+\frac{e^2}{32\pi^2}\left(\frac{2}{\sqrt{3}}\right)\frac{\eta}{f_\pi}F \tilde{F}. \label{eq:PiEtaPhoton}
\end{eqnarray}
Denoting the explicit boson-photon-photon couplings as
$G_{a\gamma\gamma}$ (see Eq.~(\ref{eq:qaLag})), $G_{\pi\gamma\gamma}$ and $G_{\eta\gamma\gamma}$
for the axion, $\pi^0$ and $\eta$ respectively, the Lagrangian above
the mass of the $\eta$ includes the following terms:
\begin{equation}
  \frac{1}{2}\left(\begin{array}{ccc} \pi^0 & \eta & a \end{array}\right) {\cal M}^2 \left(\begin{array}{c} \pi^0 \\ \eta \\ a \end{array}\right)+\frac{1}{4}\left(\begin{array}{ccc} \pi^0 & \eta & a \end{array}\right)\left(\begin{array}{c} G_{\pi\gamma\gamma} \\ G_{\eta\gamma\gamma} \\ G_{a\gamma\gamma} \end{array}\right)F\tilde{F} \label{eq:lowELag}
\end{equation}
The heavy $\eta$ and $\pi^0$ can be integrated out at low energies,
and one is left with an effective low energy coupling,
$g_{a\gamma\gamma}$, to photons involving only axions:
\begin{eqnarray}
\lefteqn{g_{a\gamma\gamma} = G_{a\gamma\gamma}-\frac{m_{\eta a}^2}{m_\eta^2}G_{\eta\gamma\gamma}} & & \nonumber \\
 & & -\frac{m_\eta^2m_{\pi^0 a}^2-m_{\pi^0 \eta}^2m_{\pi a}^2}{m_{\pi^0}^2m_\eta^2-m_{\pi^0\eta}^4}\left(G_{\pi\gamma\gamma}-\frac{m_{\pi^0\eta}^2}{m_\eta^2}G_{\eta\gamma\gamma}\right). \label{eq:agammaForm}
\end{eqnarray}
Here, we use the same normalization as reference \cite{Sikivie:1985yu} (and the ADMX collaboration), defining the coefficient of $aF\tilde{F}$ in the Lagrangian as $\frac{1}{4} g_{a\gamma\gamma}$.
The full expression for the coupling is given in
Eq.~(\ref{eq:gfull}). As a function of $z$, $w$, $E/N$, $f_a$ and
$\delta$, it can be rewritten as a function of only $z$, $f_a$ and $E/N$,
using Eq.~(\ref{eq:DeltaWMu}). Evaluating
Eq.~(\ref{eq:gfull}) when $\delta=0$, we recover the well-known axion
coupling at low energies (see {\it e.g.}~\cite{Raffelt:1990yz}):
\begin{equation}
\left.g_{a\gamma\gamma}\right|_{\delta=0} = \frac{\alpha}{2\pi}\frac{1}{f_a}\left(\frac{E}{N}-\frac{2}{3}\frac{4+z+w}{1+z+w}\right) \label{eq:GaxionF}
\end{equation}

Most experimental searches for axions make kinematic assumptions which
place direct bounds on the axion mass $m_a$ rather than $f_a$. Taking
the eigenvalues of the matrix Eq.~(\ref{eq:PiEtaAmass}), one finds a
relation between the mass and coupling constant of the pion and those
of the axions:
\begin{equation}
m_a^2 f_a^2 = m_{\pi^0}^2 f_\pi^2 F(z,w,\delta). \label{eq:massrel}
\end{equation}
Here, $m_{\pi^0}^2$ is the $\pi^0\pi^0$ entry in the neutral meson mass matrix given in Eq.~(\ref{eq:PiEtaAmass}), and the full form of $F$ is given in Eq.~(\ref{eq:Ffunct}). With
$\delta=0$, $F$ agrees with the well-known result (again, see
\cite{Raffelt:1990yz}):
\begin{equation}
\left.F\right|_{\delta=0}= \frac{z}{(1+z)(1+z+w)}. \label{eq:piAmass}
\end{equation}
Expressing $\delta$ and $w$ as functions of $z$, we find
that $F$ does not have a large numerical difference from
Eq.~(\ref{eq:piAmass}) over the allowed range of $z$.
A useful rule of thumb is that, when $\delta =0$,
\begin{equation}
m_a \approx 6~\mu\mbox{eV} \left(\frac{10^{12}~\mbox{GeV}}{f_a}\right) \label{eq:approxmass}
\end{equation}

Thus, we come to the final form for the axion-photon coupling,
expressed in terms of the axion mass, pion mass, pion decay constant,
$z$, and fundamental constants (see Eqs.~(\ref{eq:Ffunct}) and
(\ref{eq:gfull}) for the exact expression):
\begin{eqnarray}
g_{a\gamma\gamma} & = & \frac{\alpha}{2\pi}\frac{m_a}{f_\pi m_{\pi^0}}\sqrt{\frac{(1+z)(1+z+w)}{z}} \times \label{eq:Gformula} \\
& & \left(\frac{E}{N}-\frac{2}{3}\frac{4+z+w}{1+z+w}+{\cal O}(w\delta) \right) \nonumber
\end{eqnarray}
Systematic errors in the bracketed expression are introduced by two sources: the $3\%$ error caused by using a leading-order expansion in $w$ for meson masses when solving for $w$ and $\delta$, and the $10\%$ discrepancy between our function of $w$ and that in reference \cite{Kaplan:1986ru} due to our omission of additional higher order terms beyond Eq.~(\ref{eq:chiralmass2}) (see appendix \ref{sec:delta}). However, as shown in appendix \ref{sec:2flavor}, the  precise values of $w$ and $\delta$ are of lesser importance than that of $z$ in Eq.~(\ref{eq:Gformula}), and as a consequence the systematic errors are numerically negligible in the final result.

Surprisingly, for a wide range of $z$ in Eq.~(\ref{eq:zlimits2}), the exact form of $g_{a\gamma\gamma}$ (in terms of the
Kaplan-Manohar coefficient $\delta$) agrees very well with the
expression with $\delta$ set to zero. This is shown in Fig.~\ref{fig:Gcomp}. As explained in appendix B, this occurs because $\delta$ is associated with a factor of $m_s^{-1}$, and in the two flavor chiral Lagrangian both terms in Eq.~(\ref{eq:chiralmass2}) transform identically under the axion rotation Eq.~(\ref{eq:qrotation}). All this is to say that the main contribution of the
Kaplan-Manohar ambiguity to axion phenomenonology is simply to allow $z$ to vary widely; it's
inclusion was not necessary when deriving axion-pion
relations. However, this realization was only apparent in hindsight. This observation justifies the discussion in reference \cite{Moroi:1998qs}.

\begin{figure}[h]
\centering
\includegraphics[width=\columnwidth]{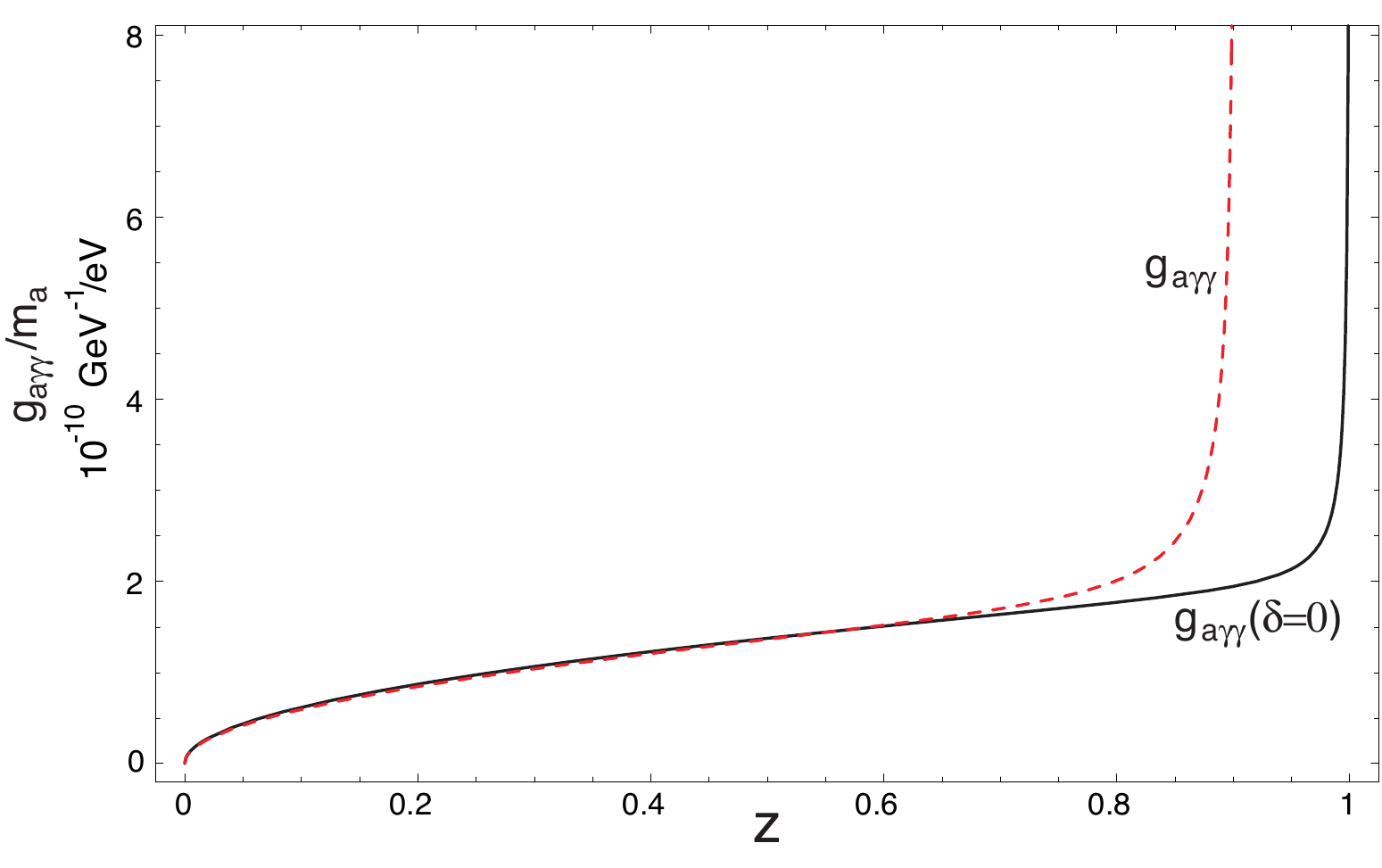}

\caption{$g_{a\gamma\gamma}/m_a$ and
  $\left.g_{a\gamma\gamma}/m_a\right|_{\delta=0}$ as a function of $z$
  for $E/N=8/3$. For the experimentally allowed range of $z$ ($0.3-0.6$ or $0.2-0.7$) both
  calculations give very similar results} \label{fig:Gcomp}
\end{figure}

An important point is the relative minus sign between the $E/N$
contribution to $g_{a\gamma\gamma}$, which has its origin in the
quantum numbers of the fermions charged under Peccei-Quinn, and the
terms involving $z$, $w$, and $\delta$, which arise from the
axion-meson mixing. For appropriate values of $E/N$, near perfect
cancelation can occur between the two contributions, strongly
suppressing the coupling. Using the Weinberg value for $z$, one finds
$-\frac{2}{3}\frac{4+z+w}{1+z+w} \approx -1.92$ with an uncertainty of $\pm 0.08$, sparking much
interest in axion models with $E/N=2$. Using the range of $z$
from Eq.~(\ref{eq:zlimits}), this term (depending on $z$, $w$, $\delta$),
is $-2.20$ at $z=0.3$ and $-1.89$ at $z=0.6$. Should the more conservative estimate of Eq.~(\ref{eq:zlimits2}) be taken, this term can vary from $-2.33$ to $-1.81$. 

What are possible values for $E/N$? The original KSVZ model used 
new heavy uncharged `quarks' as the PQ fermions $\psi$. Hence, this
model has $E/N=0$. This is sometimes referred to as `the' KSVZ model,
but it should be emphasized that the model would not suffer intrinsic
flaws if the new particles were charged. If we assume grand
unification such that the PQ fermions are in a complete multiplet of
$SU(5)$ (which occurs in DFSZ axion models
\cite{Dine:1981rt}\cite{Zhitnitsky:1980tq}) then $E/N=8/3$, as can be
quickly verified by explicit calculation of the ${\bf 5}$
representation of $SU(5)$. Other values of $E/N$ are certainly
possible, requiring only novel quantum number assignment.

At $z=0$, $-\frac{2}{3}\frac{4+z+w}{1+z+w}=8/3$, and so the coupling
to photons would vanish for GUT axion models with a massless up
quark. However, with $z=0$, the up quark can undergo a chiral rotation
as in Eq.~(\ref{eq:qrotation}) without introducing a phase to the
quark mass matrix $M$. Thus, when $z=0$, the axion is not necessary to
solve the strong CP problem.

\section{Experimental Bounds \label{sec:exp}}

With the results of the previous section, we can now examine the
recent bounds placed on the axion-photon coupling. We start with
general arguments limiting $f_a$, then consider the ADMX
results. In the construction of the Peccei-Quinn resolution to the
$\theta$ problem, there were no {\it a priori} assumptions about the
scale of $f_a$. Indeed, the original formulation of the axion model
identified the scalar $\sigma$ as the Higgs field, and thus $f_a$ as
the electroweak vev \cite{Peccei:1986pn}\cite{Krauss:1986wx}, though
this possibility was soon ruled
out \cite{Kim:1986ax}\cite{Cheng:1987gp}. General
astrophysical and cosmological considerations greatly reduce the
possible parameter space for $f_a$ and $m_a$ \footnote{For an overview
  on constraints on the axion decay constant, see
  reference \cite{Murayama:1998jb}. A more detailed review of astrophysical and
  cosmological constraints can be found in reference
  \cite{Raffelt:1990yz}}. Note that none of these constraints have
been reevaluated in light of the full uncertainty on $z$. A full
consideration of the Kaplan-Manohar ambiguity's effect on axion
physics is forthcoming \cite{longpaper}.

Axions would be produced in the early universe, and so their
properties must be such that they do not contribute a mass density
$\Omega_a > {\cal O}(1)$. There are two separate situations that must
be considered. If the universe never reached a temperature $T>f_a$
after inflation, then axions are produced via coherent oscillations due to some initial misalignment of the axion field away from the vacuum. In this scenario, axions were never in thermal equilibrium with the
rest of the matter in the universe, and so started their existence as
cold dark matter. To avoid overclosing the universe, $f_a \lesssim 10^{12}$~GeV,  translating to $m_a \gtrsim
1~\mu$eV \cite{KolbTurner}\cite{Preskill:1982cy}\cite{Abbott:1982af}\cite{Dine:1982ah}\cite{Turner:1985si}. Thus, $m_a \sim \mu$eV is very appealing, as they provide a source of cold dark matter with the correct ${\cal O}(1)$ density.

Note that this assumes a `natural' size for the initial misalignment
angle after inflation. Various arguments have been presented to avoid this bound: anthropics may make it more likely for observers to exist in a universe with an `unnaturally'  small initial misalignment angle \cite{Linde:1987bx}, some strong dynamics at high energies could force the initial misalignment to be small \cite{Dvali:1995ce}, or late entropy production from particle decays could sufficiently dilute the axions to avoid overclosure \cite{Kawasaki:1995vt}.

Alternatively, if the reheating temperature after inflation was greater than $f_a$,
then topological defects would form as the universe cooled. These
cosmic strings would radiate axions, again these relics cannot
overclose the universe. Limits from this scenario imply $f_a \lesssim
10^{11}$~GeV, or $m_a \gtrsim 10^{-5}$~eV \cite
{Davis:1989nj}\cite{Harari:1987ht}. Note that models with $N\geq 2$ have an exact $Z_N$ symmetry that is spontaneously broken, and hence lead to cosmologically unacceptable domain walls.

In addition to these cosmological arguments, axion properties can be bounded by several additional considerations. These fall into two broad categories: direct searches and limits on novel energy-loss mechanisms from stars. These constrain the axion decay constant on the scale $f_a <10^9$~GeV ($m_a>0.01$~eV), and so are not relevant for considerations of axionic cold dark matter. For a recent review of these constraints, see reference \cite{Raffelt:2006cw}. We note that the CAST collaboration has released new results \cite{Andriamonje:2007ew} which supersede those quoted in reference \cite{Raffelt:2006cw}.

With this in mind, the ADMX experiment is very interesting
\cite{Asztalos:2001tf}\cite{Duffy:2006aa}\cite{Asztalos:2003px}. ADMX is a laboratory
experiment searching for axions in the galactic dark matter halo
through their coupling to photons. Axions interact with a strong
magnetic field, allowing them to transition to photons with energy
equal to the rest mass $m_a$ plus small kinetic corrections. The
conversion occurs within a resonant cavity, which can be `tuned' to
look for photons of a particular wavelength \cite{Sikivie:1983ip}, corresponding to an axion
mass between $1.9$ and $3.3~\mu$eV for the ADMX apperatus \cite{Asztalos:2003px}.

\begin{figure}[h]
\centering
\includegraphics[width=\columnwidth]{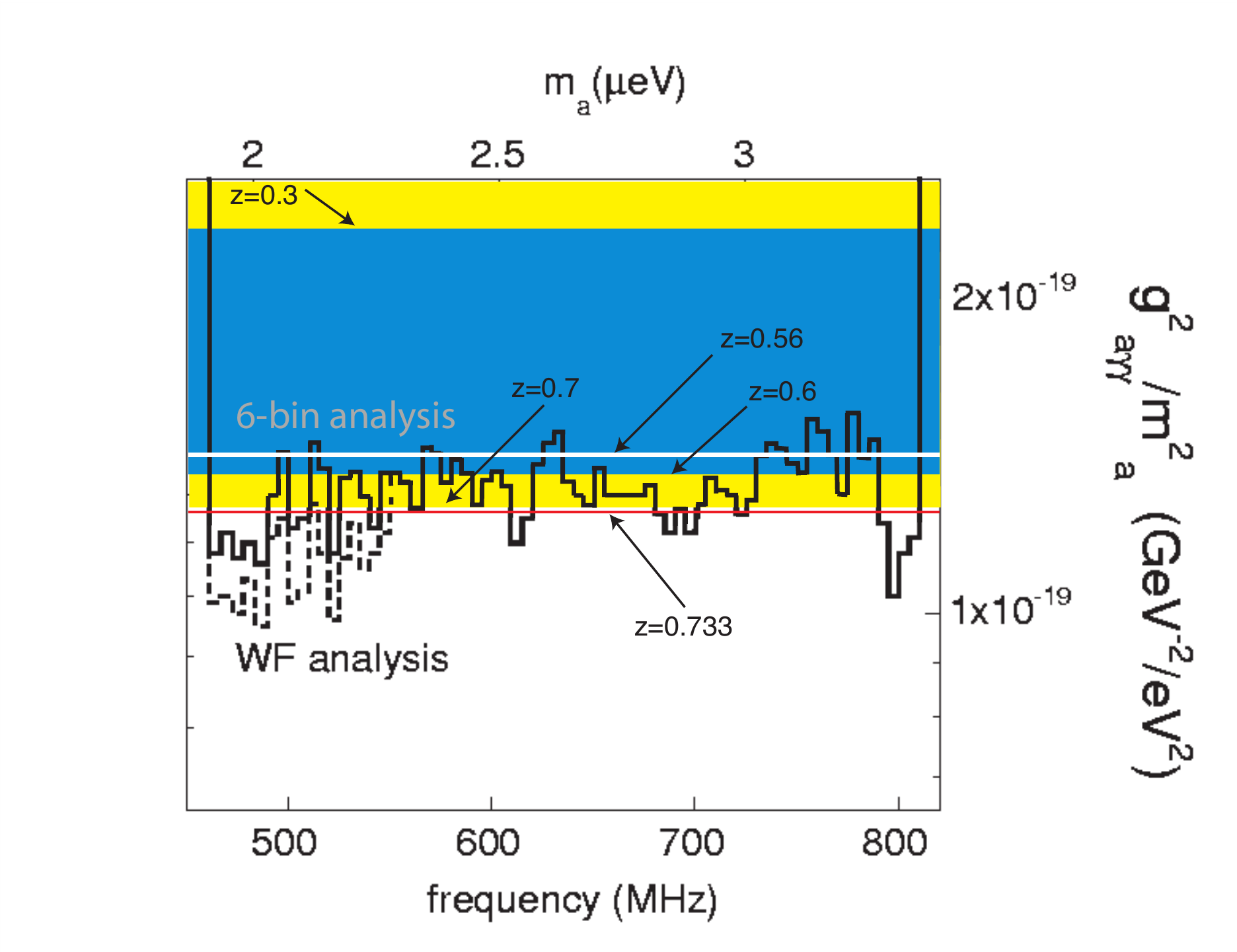}

\caption{The experimentally excluded region at 90\% confidence from the ADMX experiment, taken from \cite{Asztalos:2003px}. Our estimate for the experimental upper bound for all values of $m_a$, $g_{a\gamma\gamma}^2/m_a^2 \leq 1.44 \times 10^{-19}$~GeV$^{-2}$/eV$^2$, corresponds to the coupling evaluated at the Weinberg value $z=0.56$. The theoretically predicted values of $(g_{a\gamma\gamma}/m_a)^2$ for various values of $z$ are indicated. The $z$ value of $0.733$ corresponds to the smallest coupling for $E/N$. The values for $z \lesssim 0.3$ lie above the plot range.  \label{fig:ADMX}}.
\end{figure}

ADMX used a local dark matter halo density of $0.45$~GeV/cm$^3$; assuming that this 
is composed entirely of axions, a lack
of detected signal can be interpreted as an upper limit on the
axion-to-photon coupling $g_{a\gamma\gamma}$. Alternatively, one could
place limits on the percentage of the halo which is made of axions. The excluded region is nearly
constant if plotted as $g_{a\gamma\gamma}/m_a$ versus $m_a$
\cite{Asztalos:2001tf}.  However, there is considerable structure on small scales, as seen in Fig.~\ref{fig:ADMX} \cite{Asztalos:2003px}. Furthermore, more stringent bounds are available from ADMX's high resolution search over the narrow mass range $1.98-2.17$~eV \cite{Duffy:2006aa}.

Using the Weinberg value for $z$ and assigning $E/N=0$ for KSVZ and
$8/3$ for DSFZ, the ADMX collaboration compared the experimental upper
limit to the predicted coupling for these two common axion models. In
doing so, ADMX ruled out the KSVZ model as the sole source of dark matter in the galaxy
(at 90\% confidence) in the mass range over which they
scanned (see Fig.~\ref{fig:ADMX}). However, we note that this exclusion occurs on the margins of the experimental limits \footnote{We discovered a 5\% mismatch in ($g_{a\gamma\gamma}/m_a)^2$ between the ADMX value and our analysis. This seems to be the result of different input parameters for $f_\pi$ and $m_\pi$ in Eq.~(\ref{eq:piAmass}). Fig.~\ref{fig:ADMX} shows our values for the coupling.}.  

In addition, there are three additional
considerations in play. First, $z$ has a larger range of possible values
than previously considered.  Secondly, KSVZ models can have $E/N
\neq 0$. Lastly, the precise value of the local galactic halo density is uncertain, and may not be composed of only axions. The ADMX limit can therefore be considered an exclusion of a `benchmark' KSVZ model, with $E/N=0$, together with the extra assumptions that $z=0.56\pm 0.05$, and $\rho_{DM} = 0.45$~GeV/cm$^3$. 

\begin{figure}[h]
\centering
\includegraphics[width=\columnwidth]{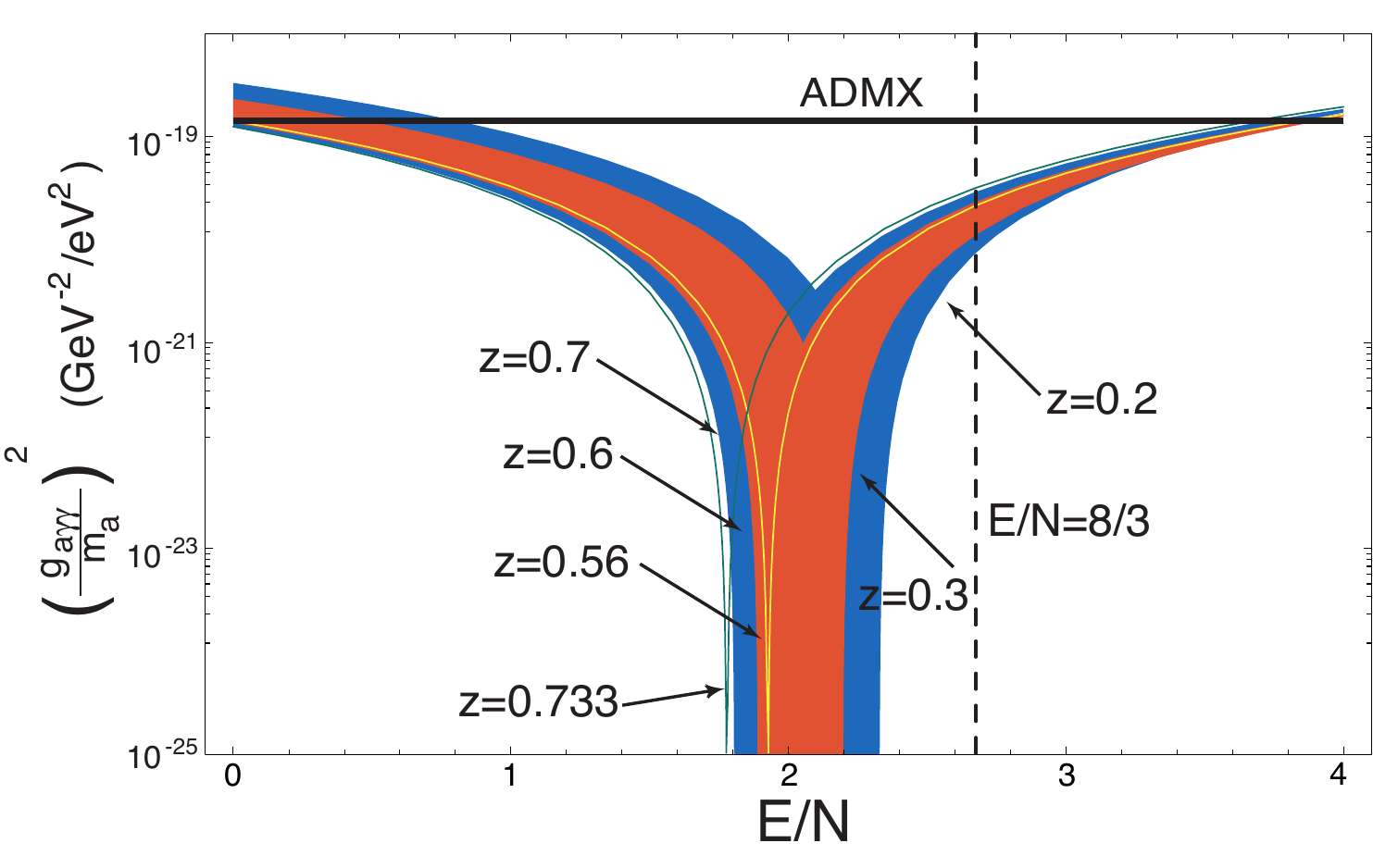}

\caption{$(g_{a\gamma\gamma}/m_a)^2$ versus $E/N$ for varying values
  of $z$. The ADMX exclusion line is the exclusion line as in Fig.~\ref{fig:ADMX}, and corresponds to $z=0.56$ and $E/N=0$. The value of $(g_{a\gamma\gamma}/m_a)^2$ as a function of $E/N$ for the
  Weinberg value, $z=0.56$, is indicated. The orange band
  corresponds to $0.3<z<0.6$, while the wider blue band is the range from $0.2<z<0.7$. The minimal coupling at $E/N=0$, corresponding to $z=0.733$, is also shown. \label{fig:exc}}
\end{figure}

In Fig.~\ref{fig:exc} the $E/N$ dependence of
$(g_{a\gamma\gamma}/m_a)^2$ is plotted for the full range of allowed
$z$. The coupling is divided by $m_a$ so that the experimental limit
is independent of $m_a$, and squared to agree with the ADMX
convention. For $E/N=0$, the allowed range of $z$ permits only a small 
reduction in the axion-photon coupling. However, as is more clearly demonstrated in 
Fig.~\ref{fig:ADMX}, for $z=0.6$ this reduction is sufficient to allow a small window in $(g_{a\gamma\gamma}/m_a)^2$
which has not yet been ruled out by ADMX. 

\begin{figure}[h]
\centering
\includegraphics[width=\columnwidth]{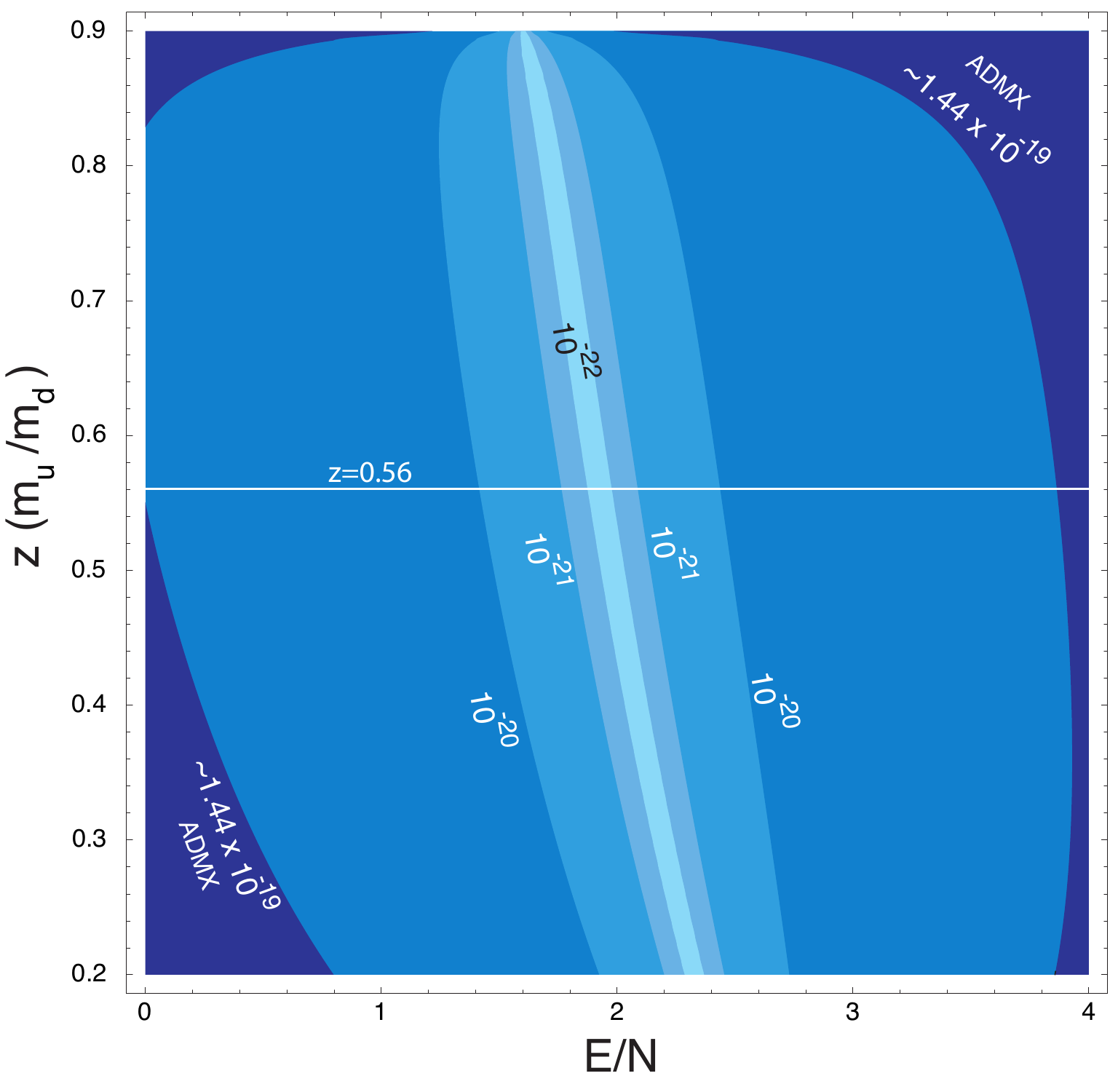}

\caption{Contour plot of $(g_{a\gamma\gamma}/m_a)^2$ in units of
  GeV$^{-2}$/eV$^2$ as a function of $z$ and $E/N$. Darker regions
  indicate larger values. The line marked ``ADMX" is the experimental exclusion estimated from  90\% CL exclusion in Fig.~\ref{fig:ADMX}, taken from \cite{Asztalos:2003px}. \label{fig:contour}}.
\end{figure}

An alternate visualization is shown in Fig.~\ref{fig:contour}, which demonstrates
the large parameter space of $E/N$ versus $z$ which has not
been excluded by ADMX. As can be
seen, even with the Weinberg value of $z$, a sizable region is still experimentally allowed. Additionally, when $E/N=0$, values of the light quark mass ratio $0.56 < z \lesssim 0.8$ allow $g_{a\gamma\gamma}$ to escape the experimental bound. Thus, even with the more aggressive bound on $z$ from Eq.~(\ref{eq:zlimits}), it is premature for us to conclude that the KSVZ model in the $2-3~\mu$eV range cannot be  the dominant source of dark matter in the galaxy.

\section{Conclusions \label{sec:conc}}

The $\mu$eV axion mass scale probed by ADMX is especially interesting,
as such axions would provide an ideal candidate for dark matter. In
order to compare experimental results with theory, an accurate
knowledge of pion physics and quark masses is required. As we have
demonstrated, the Kaplan-Manohar ambiguity in quark mass implies that
the canonical Weinberg value of $z$, used in the derivation of
axion-photon couplings, is very suspect.

The ratio $z$ is poorly constrained from QCD physics, significantly  
expanding the theoretical uncertainty on the photon coupling for
specific models. Allowing $z$ to vary within the range $0.3-0.6$ $(0.2-0.7)$ allows the axion coupling to photons due to meson mixing to vary by $10\%$ ($15\%$). This dwarfs the systematic error in our calculations due to additional higher order terms in the chiral Lagrangian.

Due to this uncertainty, even with a
fixed value of $E/N=0$, the KSVZ model can escape the ADMX experimental
bound, as can be seen in Fig.~\ref{fig:exc}. Similarly, outside the range $E/N \approx 2.1\pm0.2$, the
possible variation in the value of $z$ generally lowers the measured
combination $(g_{a\gamma\gamma}/m_a)^2$ by factors of a few.

In the region $E/N \approx 2.1\pm0.2$, complete cancelation between
the $E/N$ and the axion-meson mixing term allows the magnitude of
$g_{a\gamma\gamma}$ to evade experimental bounds by many orders of
magnitude, even reaching zero in some cases. Perfect cancelation
requires fine-tuning, but it is notable that near perfect
cancelation occurs for a considerable range of $E/N$ and $z$ values,
as can be seen in Figs.~\ref{fig:exc} and \ref{fig:contour}.

The DFSZ model is essentially unconstrained by the ADMX results. As can be clearly seen in Figs.~\ref{fig:exc} and \ref{fig:contour}, even with the Weinberg value for $z$, models with $E/N=8/3$ fall well below the experimental limit. Fortunately for the discovery prospects, the GUT value of $8/3$
can only be cancelled by $z=0$.  In addition to being outside the range we consider, this
value for $z$ would entirely negate the need for the PQ mechanism as a
solution to the strong CP problem. Therefore, one should remain optimistic for the prospects for future bounds on the DFSZ model.

It should also be noted that  the ADMX limit excludes only a benchmark KSVZ model with $E/N=0$. Naively, one might assume that a model with no explicit
coupling between the axions and photons due to charged PQ fermions
would be the most conservative choice when placing experimental
bounds. However, as has been demonstrated, the smallest couplings to
photons instead occur when $E/N\approx 2$, as a result of cancelations
in opposite sign chiral rotations on the PQ fermions and standard
model quarks.

Should the ADMX collaboration increase their sensitivity by at least an order of magnitude 
as projected \cite{Carosi:2007uc},
then axionic galactic dark matter models with either $E/N=0$ or $E/N=8/3$ would be 
experimentally accessible for all possible values of $z$. For both of these
points, near-perfect cancellation cannot occur for the range of
$z$. 

As stated in section~\ref{sec:exp}, there are numerous other axion
bounds besides ADMX. The large uncertainty in $z$ from the
Kaplan-Manohar term should have similar effects on these limits
as well. Preliminary work indicates that the bounds should in general
be loosened by factors of ${\cal O}(1)$. As many astrophysical
constraints depend on coupling to nucleons and electrons, which
generically lack the possibility of cancelation (as with the $E/N$
term), one would not expect regions of parameter space where the
bounds can be reduced by many orders of magnitude. ADMX, in presenting
the strongest limits on axion phenomenonology, is also the experiment
where the uncertainty in $z$ is most critical.

\appendix

\section{Mass Matrices \label{sec:delta}}

Expanding out Eq.~(\ref{eq:chiralmass2}) and taking the terms second
order in the meson and axion fields, one can determine the NG boson
mass matrix and the axion-$\pi^0$-$\eta$ mass squared matrix ${\cal
  M}^2$. The elements of this matrix are
\begin{eqnarray}
m_{\pi^0}^2 & = & \mu m_u \frac{(1+z)(1+\delta)}{z} \nonumber \\
m_{\pi^0 \eta}^2 & = & \mu m_u  \frac{(1-z)(\delta-1)}{\sqrt{3}z} \nonumber \\
m_{\eta}^2 & = & \mu m_u  \frac{(w+4z+wz+w(1+4w+z)\delta)}{3wz} \nonumber \\
m_{\pi^0 a}^2 & = & \mu m_u \frac{(1-z)(1+2\delta)f_\pi}{3zf_a} \label{eq:PiEtaAmass} \\
m_{\eta a}^2 & = & \mu m_u  \frac{(2z-w-wz+2w(1-2w+z)\delta)f_\pi}{3\sqrt{3} w z f_a} \nonumber \\
m_{a}^2 & = &  \mu m_u  \frac{(w+z+wz+4w(1+w+z)\delta)f_\pi^2}{9wzf_a^2} \nonumber
\end{eqnarray}
Note that these are the entries in the mass matrix, not the mass
eigenvalues of the physical particles. In addition, the masses of the
other NG bosons are
\begin{eqnarray}
(m_{\pi^\pm}^2)_{\mbox{\small QCD}} & = &  \mu m_u  \frac{(1+z)(1+\delta)}{z} \nonumber \\
m_{K^0}^2 = m_{\bar{K}^0}^2 & = & \mu m_u \frac{(w+z)(1+w\delta)}{wz} \label{eq:Kmass} \\
(m_{K^\pm}^2)_{\mbox{\small QCD}} & = &  \mu m_u \frac{(1+w)(z+w\delta)}{wz} \nonumber
\end{eqnarray}
The subscript QCD refers to the mass squared ignoring electromagnetic contributions. To leading order, the EM contributions to the charged pion and kaon are the same \footnote{Higher order effects cause the EM contributions to $\pi^\pm$ and $K^\pm$ to differ by ${\cal O}(80\%)$ \cite{Donoghue:1993hj}. However, this simply has the effect of changing the value of $z$ derived from the chiral Lagrangian. For our purposes, $z$ is a free parameter, and so we can neglect the higher order contributions to the pion/kaon mass difference.}.

To determine $\delta$, $w$, and $\mu m_u$ as functions of $z$, three physical observables involving the meson masses and mixings Eqs.~(\ref{eq:PiEtaAmass}) and (\ref{eq:Kmass}) are constructed. The physical $\pi^0$ mass can be extracted from the mixing of $\pi^0$ and $\eta$:
\begin{eqnarray}
\lefteqn{(m_{\pi^0}^2)_{\mbox{\small phys.}}  =  m_{\pi^0}^2-\frac{m_{\pi^0\eta}^4}{m_\eta^2-m_{\pi^0}^2} } & &  \label{eq:Pi0} \\
 & = &  \mu m_u \left(\frac{(1+z)(1+\delta)}{z}-\frac{w(z-1)^2(\delta-1)^2}{4z^2}\right)+{\cal O}(w^2) \nonumber
\end{eqnarray}
The electromagnetic contributions to the physical $\pi^\pm$ and $K^\pm$ mesons can be eliminated at leading order by taking the mass difference:
\begin{equation}
(m_{K^\pm}^2-m_{\pi^\pm}^2)_{\mbox{\small phys.}}  =  \mu m_u\left(\frac{1}{w}-\frac{1}{z}-\delta\right)+{\cal O}(w) \label{eq:ChargeDiff}
\end{equation}
For the third observable, we expand the $K^0$ mass to leading order in $w$:
\begin{equation}
m_{K^0}^2 =  \mu m_u \left(\frac{1}{w}+\frac{1}{z}+\delta \right)+{\cal O}(w)\label{eq:K0mass}.
\end{equation}
For calculational purposes, we truncated Eqs.~(\ref{eq:Pi0})-(\ref{eq:K0mass}) to the indicated order in $w$. However, using the exact expressions only introduces an error of $<3\%$ in the numerical solutions for $w$, $\delta$, and $\mu m_u$.

Combining Eqs.~(\ref{eq:Pi0}), (\ref{eq:ChargeDiff}), and (\ref{eq:K0mass}), we solve for $\mu m_u$, $w$, and $\delta$ in terms of $z$ and the observable masses of $\pi^0$, $K^0$, and the charged meson mass difference. Two sets of solutions are obtained, one gives negative values for $w$, and so is discarded. Using the known meson masses, the remaining set of solutions is
\begin{eqnarray}
\delta(z) & = & \frac{1.78(-0.56+z)}{-1.78+z} \nonumber \\
w(z) & = & \frac{0.0279(z-1.78)z}{z^2-1} \label{eq:DeltaWMu} \\
\mu m_u(z) & = & (81.1~\mbox{MeV})^2 \frac{(z-1.78)z}{z^2-1} \nonumber
\end{eqnarray}
Fig.~\ref{fig:delta} shows $\delta$ as a function of $z$. Comparison of our function $w(z)$ with that given in reference \cite{Kaplan:1986ru} reveals $\sim 10\%$ discrepancy, originating in our neglect of additional higher order terms in the chiral Lagrangian.

Diagonalizing the neutral meson mass squared matrix ${\cal M}^2$
allows one to find the physical masses for the $\pi^0$, $\eta$, and
axion $a$. Writing the axion mass in terms of the pion mass, we find
the functional form of $F(z,w,\delta)$, introduced in
Eq.~(\ref{eq:massrel}).
\begin{eqnarray}
\lefteqn{F(z,w,\delta) =  \left\{\left[z+(w+z+wz)\delta\right]^2+4zw^2\delta^3 \right\}} && \nonumber \\
 &  & / \left\{(1+z)(1+\delta)[(z+z^2+w\delta)(1+w+w\delta)\right. \nonumber \\
 & & \left. +z(\delta+z\delta-zw-w\delta+w^2\delta+w\delta^2+w^2\delta^2)] \right\}. \label{eq:Ffunct}
\end{eqnarray}
Recall that $w$ and $\delta$ are functions of $z$. 

As in Eq.~(\ref{eq:lowELag}), diagonalizing the mass matrix also
determines the low-energy axion-photon coupling $g_{a\gamma\gamma}$ in
terms of the explicit couplings of $a$, $\pi^0$ and $\eta$ to photons,
as given in Eq.~(\ref{eq:agammaForm}). Using the explicit mass squared
terms from Eq.~(\ref{eq:PiEtaAmass}), the full form of
$g_{a\gamma\gamma}$ is found to be
\begin{eqnarray}
g_{a\gamma\gamma} & = & \frac{\alpha}{2\pi}\frac{1}{f_a}\left(\frac{E}{N}-\frac{2}{3}C\right) \label{eq:gfull} \\
C & \equiv & \{z(4+w+z)+[w^2(1+z)+z(4+z)+ \nonumber \\
 & & w(4+z^2)]\delta-2w[w(z-2)-2z]\delta^2 \} \nonumber \\
 & & /\{(z+z^2+w\delta)(1+w+w\delta) \nonumber \\
 & & +z(\delta+z\delta-zw-w\delta+w^2\delta+w\delta^2+w^2\delta^2)\} \nonumber
 \end{eqnarray}

\section{Two Flavor Analysis \label{sec:2flavor}}

When working with only two quark flavors, including the axion proceeds identically up to Eq.~(\ref{eq:qrotation}). At this point, to avoid kinetic mixing terms in the chiral Lagrangian, the up and down quark fields are rotated by
\begin{equation}
q \to \exp\left(\frac{ia}{f_a}\frac{\gamma_5}{4}\right) q \label{eq:2qrotation}
\end{equation}
as there are only two quark flavors being rotated (four fields altogether, once handedness is included).
 Following the three-flavor analysis, we arrive at Eq.~(\ref{eq:chiralmass2}), with $M = \mbox{diag}(m_u,m_d)$ and axions included by rotating $M$ by $\exp(ia/2f_a)$ (twice the rotation in Eq.~(\ref{eq:2qrotation})).

At this point, it is clear why $\delta$ enters into the axion phenomenonology only in the form $w\delta$. Naively, one would expect $g_{a\gamma\gamma}$ to have terms of order $\delta$, $w\delta$, $\delta^2$, {\it etc.} The two flavor analysis corresponds formally to sending $w\to 0$, so the leading corrections would be of the form $\delta$. However, the leading order term in Eq.~(\ref{eq:chiralmass2}) transforms as $e^{ia/2f_a}$, while the $(\det(M)M^{-1})^\dag$ terms transforms as $e^{-ia/f_a}e^{ia/2f_a}=e^{-ia/2f_a}$. Including the hermitian conjugates, we find that both the leading order piece and the Kaplan-Manohar correction both transform identically under the axion rotation. 

Therefore, the Kaplan-Manohar term can be considered as a simple redefinition of $m_u$ and $m_d$ as far as the axion is concerned. Thus, there is no ${\cal O}(\delta)$ correction to the axion couplings. The effect to the higher order term can be completely parametrized in terms of $z$. 

\begin{acknowledgments}
  This work was supported in part by the U.S. DOE
  under Contract DE-AC03-76SF00098, and in part by the NSF
  under grant PHY-04-57315.
  
  The authors would also like to thank Aneesh Manohar, Georg Raffelt, and Leslie Rosenberg for their advice and comments in the preparation of this work.
\end{acknowledgments}


\begin{thebibliography}{99}

\bibitem{Kamionkowski:1997zb}
  M.~Kamionkowski,
  %``WIMP and axion dark matter,''
  arXiv:hep-ph/9710467.
  %%CITATION = HEP-PH/9710467;%%

\bibitem{Weinberg:1975ui}
  S.~Weinberg,
  %``The U(1) Problem,''
  Phys.\ Rev.\  D {\bf 11}, 3583 (1975).
  %%CITATION = PHRVA,D11,3583;%%
  
\bibitem{Altarev:1981zp}
  I.~S.~Altarev {\it et al.},
  %``A New Upper Limit On The Electric Dipole Moment Of The Neutron,''
  Phys.\ Lett.\  B {\bf 102}, 13 (1981).
  %%CITATION = PHLTA,B102,13;%%

\bibitem{Pendlebury1984}
 J.~M.~Pendlebury {\it et al.},
 %``Search for a neutron electric dipole moment,''
 Phys.\ Lett.\ B {\bf 136}, 327 (1984).

\bibitem{Baker:2006ts}
  C.~A.~Baker {\it et al.},
  %``An improved experimental limit on the electric dipole moment of the
  %neutron,''
  Phys.\ Rev.\ Lett.\  {\bf 97}, 131801 (2006)
  [arXiv:hep-ex/0602020].
  %%CITATION = PRLTA,97,131801;%%

\bibitem{Peccei:1977ur}
  R.~D.~Peccei and H.~R.~Quinn,
  %``Constraints Imposed By CP Conservation In The Presence Of Instantons,''
  Phys.\ Rev.\  D {\bf 16}, 1791 (1977).
  %%CITATION = PHRVA,D16,1791;%%

\bibitem{Peccei:1977hh}
  R.~D.~Peccei and H.~R.~Quinn,
  %``CP Conservation In The Presence Of Instantons,''
  Phys.\ Rev.\ Lett.\  {\bf 38}, 1440 (1977).
  %%CITATION = PRLTA,38,1440;%%
  
\bibitem{Weinberg:1977ma}
  S.~Weinberg,
  %``A New Light Boson?,''
  Phys.\ Rev.\ Lett.\  {\bf 40}, 223 (1978).
  %%CITATION = PRLTA,40,223;%%

\bibitem{Wilczek:1977pj}
  F.~Wilczek,
  %``Problem Of Strong P And T Invariance In The Presence Of Instantons,''
  Phys.\ Rev.\ Lett.\  {\bf 40}, 279 (1978).
  %%CITATION = PRLTA,40,279;%%
  
\bibitem{Raffelt:2006cw}
  G.~G.~Raffelt,
  %``Astrophysical axion bounds,''
  arXiv:hep-ph/0611350.
  %%CITATION = HEP-PH/0611350;%%

\bibitem{Zavattini:2005tm}
  E.~Zavattini {\it et al.}  [PVLAS Collaboration],
  %``Experimental observation of optical rotation generated in vacuum by a
  %magnetic field,''
  Phys.\ Rev.\ Lett.\  {\bf 96}, 110406 (2006)
  [arXiv:hep-ex/0507107].
  %%CITATION = PRLTA,96,110406;%%

\bibitem{Maiani:1986md}
  L.~Maiani, R.~Petronzio and E.~Zavattini,
  %``EFFECTS OF NEARLY MASSLESS, SPIN ZERO PARTICLES ON LIGHT PROPAGATION IN A
  %MAGNETIC FIELD,''
  Phys.\ Lett.\  B {\bf 175}, 359 (1986).
  %%CITATION = PHLTA,B175,359;%%

\bibitem{Raffelt:1988}
  G.~G.~Raffelt and L.~Stodolsky,
  %"Mixing of the photon with low-mass particles,"
  Phys.\ Rev.\ D {\bf 37}, 1237 (1988)
  
\bibitem{Asztalos:2001tf}
  S.~Asztalos {\it et al.},
  %``Large-scale microwave cavity search for dark-matter axions,''
  Phys.\ Rev.\  D {\bf 64}, 092003 (2001).
  %%CITATION = PHRVA,D64,092003;%%

\bibitem{Kim:1979if}
  J.~E.~Kim,
  %``Weak Interaction Singlet And Strong CP Invariance,''
  Phys.\ Rev.\ Lett.\  {\bf 43}, 103 (1979).
  %%CITATION = PRLTA,43,103;%%

\bibitem{Shifman:1979if}
  M.~A.~Shifman, A.~I.~Vainshtein and V.~I.~Zakharov,
  %``Can Confinement Ensure Natural CP Invariance Of Strong Interactions?,''
  Nucl.\ Phys.\  B {\bf 166}, 493 (1980).
  %%CITATION = NUPHA,B166,493;%%

\bibitem{Chang:1993gm}
  S.~Chang and K.~Choi,
  %``Hadronic axion window and the big bang nucleosynthesis,''
  Phys.\ Lett.\  B {\bf 316}, 51 (1993)
  [arXiv:hep-ph/9306216].
  %%CITATION = PHLTA,B316,51;%%

\bibitem{Moroi:1998qs}
  T.~Moroi and H.~Murayama,
  %``Axionic hot dark matter in the hadronic axion window,''
  Phys.\ Lett.\  B {\bf 440}, 69 (1998)
  [arXiv:hep-ph/9804291].
  %%CITATION = PHLTA,B440,69;%%

\bibitem{Kaplan:1986ru}
  D.~B.~Kaplan and A.~V.~Manohar,
  %``Current Mass Ratios Of The Light Quarks,''
  Phys.\ Rev.\ Lett.\  {\bf 56}, 2004 (1986).
  %%CITATION = PRLTA,56,2004;%%

\bibitem{Svrcek:2006yi}
  P.~Svr\v{c}ek and E.~Witten,
  %``Axions in string theory,''
  JHEP {\bf 0606}, 051 (2006)
  [arXiv:hep-th/0605206].
  %%CITATION = JHEPA,0606,051;%%

\bibitem{Dine:1981rt}
  M.~Dine, W.~Fischler and M.~Srednicki,
  %``A Simple Solution To The Strong CP Problem With A Harmless Axion,''
  Phys.\ Lett.\  B {\bf 104}, 199 (1981).
  %%CITATION = PHLTA,B104,199;%%

\bibitem{Zhitnitsky:1980tq}
  A.~R.~Zhitnitsky,
  %``On Possible Suppression Of The Axion Hadron Interactions. (In Russian),''
  Sov.\ J.\ Nucl.\ Phys.\  {\bf 31}, 260 (1980)
  [Yad.\ Fiz.\  {\bf 31}, 497 (1980)].
  %%CITATION = YAFIA,31,497;%%
  
\bibitem{Weinberg:1977hb}
  S.~Weinberg,
  %``The Problem Of Mass,''
  Trans.\ New York Acad.\ Sci.\  {\bf 38}, 185 (1977).
  %%CITATION = TNYAA,38,185;%%

\bibitem{Gasser:1982ap}
  J.~Gasser and H.~Leutwyler,
  %``Quark Masses,''
  Phys.\ Rept.\  {\bf 87}, 77 (1982).
  %%CITATION = PRPLC,87,77;%%

\bibitem{Georgibook}
  H.~Georgi, {\it Weak Interactions and Modern Particle Theory} (The Benjamin-Cummings Publishing Company, Inc., Menlo Park, 1984).

\bibitem{Manohar:2004fp}
   A.~V.~Manohar and C.~T.~Sachrajda,
   ``Quark Masses,''
  %\href{http://www.slac.stanford.edu/spires/find/hep/www?irn=5997500}{SPIRES entry}
  in
  W.~M.~Yao {\it et al.}  [Particle Data Group],
  %``Review of particle physics,''
  J.\ Phys.\ G {\bf 33}, 1 (2006).
  %%CITATION = JPHGB,G33,1;%%

\bibitem{Maltman:1989sx}
  K.~Maltman, T.~Goldman and G.~J.~.~Stephenson,
  %``Corrections To The Extraction Of Light Quark Mass Ratios From Pseudoscalar
  %Meson Splittings,''
  Phys.\ Lett.\  B {\bf 234}, 158 (1990).
  %%CITATION = PHLTA,B234,158;%%

\bibitem{Leutwyler:1989pn}
  H.~Leutwyler,
  %``How About M (U) = 0?,''
  Nucl.\ Phys.\  B {\bf 337}, 108 (1990).
  %%CITATION = NUPHA,B337,108;%%

\bibitem{Sikivie:1985yu}
  P.~Sikivie,
  %``Detection Rates For 'Invisible' Axion Searches,''
  Phys.\ Rev.\  D {\bf 32}, 2988 (1985)
  [Erratum-ibid.\  D {\bf 36}, 974 (1987)].
  %%CITATION = PHRVA,D32,2988;%%

\bibitem{Raffelt:1990yz}
  G.~G.~Raffelt,
  %``Astrophysical methods to constrain axions and other novel particle
  %phenomena,''
  Phys.\ Rept.\  {\bf 198}, 1 (1990).
  %%CITATION = PRPLC,198,1;%%

\bibitem{Peccei:1986pn}
  R.~D.~Peccei, T.~T.~Wu and T.~Yanagida,
  %``A Viable Axion Model,''
  Phys.\ Lett.\  B {\bf 172}, 435 (1986).
  %%CITATION = PHLTA,B172,435;%%

\bibitem{Krauss:1986wx}
  L.~M.~Krauss and F.~Wilczek,
  %``A Shortlived Axion Variant,''
  Phys.\ Lett.\  B {\bf 173}, 189 (1986).
  %%CITATION = PHLTA,B173,189;%%

\bibitem{Kim:1986ax}
  J.~E.~Kim,
  %``Light Pseudoscalars, Particle Physics and Cosmology,''
  Phys.\ Rept.\  {\bf 150}, 1 (1987).
  %%CITATION = PRPLC,150,1;%%

\bibitem{Cheng:1987gp}
  H.~Y.~Cheng,
  %``The Strong CP Problem Revisited,''
  Phys.\ Rept.\  {\bf 158}, 1 (1988).
  %%CITATION = PRPLC,158,1;%%

\bibitem{Murayama:1998jb}
  H.~Murayama, G.~G.~Raffelt, C.~Hagmann, K.~van Bibber and L.~J.~Rosenberg,
  %``Axions and other very light bosons: in Review of Particle Physics (RPP
  %1998),''
  Eur.\ Phys.\ J.\  C {\bf 3}, 264 (1998).
  %%CITATION = EPHJA,C3,264;%%

\bibitem{longpaper} M.R. Buckley and H. Murayama, in preparation.  

\bibitem{KolbTurner}
  E.~W.~Kolb and M.~S.~Turner,
  {\it The Early Universe} (Addison-Wesley Publishing Company, Redwood City, 1990).

\bibitem{Preskill:1982cy}
  J.~Preskill, M.~B.~Wise and F.~Wilczek,
  %``Cosmology of the invisible axion,''
  Phys.\ Lett.\  B {\bf 120}, 127 (1983).
  %%CITATION = PHLTA,B120,127;%%

\bibitem{Abbott:1982af}
  L.~F.~Abbott and P.~Sikivie,
  %``A cosmological bound on the invisible axion,''
  Phys.\ Lett.\  B {\bf 120}, 133 (1983).
  %%CITATION = PHLTA,B120,133;%%

\bibitem{Dine:1982ah}
  M.~Dine and W.~Fischler,
  %``The not-so-harmless axion,''
  Phys.\ Lett.\  B {\bf 120}, 137 (1983).
  %%CITATION = PHLTA,B120,137;%%
  
\bibitem{Turner:1985si}
  M.~S.~Turner,
  %``Cosmic And Local Mass Density Of Invisible Axions,''
  Phys.\ Rev.\  D {\bf 33}, 889 (1986).
  %%CITATION = PHRVA,D33,889;%%
  
\bibitem{Linde:1987bx}
  A.~D.~Linde,
  %``Inflation And Axion Cosmology,''
  Phys.\ Lett.\  B {\bf 201}, 437 (1988).
  %%CITATION = PHLTA,B201,437;%%

\bibitem{Dvali:1995ce}
  G.~R.~Dvali,
  %``Removing the cosmological bound on the axion scale,''
  arXiv:hep-ph/9505253.
  %%CITATION = HEP-PH/9505253;%%

\bibitem{Kawasaki:1995vt}
  M.~Kawasaki, T.~Moroi and T.~Yanagida,
  %``Can Decaying Particles Raise the Upperbound on the Peccei-Quinn Scale?,''
  Phys.\ Lett.\  B {\bf 383}, 313 (1996)
  [arXiv:hep-ph/9510461].
  %%CITATION = PHLTA,B383,313;%%

\bibitem{Davis:1989nj}
  R.~L.~Davis and E.~P.~S.~Shellard,
  %``Do Axions Need Inflation?,''
  Nucl.\ Phys.\  B {\bf 324}, 167 (1989).
  %%CITATION = NUPHA,B324,167;%%

\bibitem{Harari:1987ht}
  D.~Harari and P.~Sikivie,
  %``On The Evolution Of Global Strings In The Early Universe,''
  Phys.\ Lett.\  B {\bf 195}, 361 (1987).
  %%CITATION = PHLTA,B195,361;%%

\bibitem{Andriamonje:2007ew}
  S.~Andriamonje {\it et al.}  [CAST Collaboration],
  %``An improved limit on the axion-photon coupling from the CAST experiment,''
  JCAP {\bf 0702}, 010 (2007)
  [arXiv:hep-ex/0702006].
  %%CITATION = JCAPA,0702,010;%%

\bibitem{Duffy:2006aa}
  L.~D.~Duffy {\it et al.},
  %``A high resolution search for dark-matter axions,''
  Phys.\ Rev.\  D {\bf 74}, 012006 (2006)
  [arXiv:astro-ph/0603108].
  %%CITATION = PHRVA,D74,012006;%%

\bibitem{Sikivie:1983ip}
  P.~Sikivie,
  %``Experimental tests of the *invisible* axion,''
  Phys.\ Rev.\ Lett.\  {\bf 51}, 1415 (1983)
  [Erratum-ibid.\  {\bf 52}, 695 (1984)].
  %%CITATION = PRLTA,51,1415;%%
  
\bibitem{Asztalos:2003px}
  S.~J.~Asztalos {\it et al.},
  %``An improved RF cavity search for halo axions,''
  Phys.\ Rev.\  D {\bf 69}, 011101 (2004)
  [arXiv:astro-ph/0310042].
  %%CITATION = PHRVA,D69,011101;%%

\bibitem{Carosi:2007uc}
  G.~Carosi and K.~van Bibber,
  %``Cavity microwave searches for cosmological axions,''
  arXiv:hep-ex/0701025.
  %%CITATION = HEP-EX/0701025;%%
  
\bibitem{Donoghue:1993hj}
  J.~F.~Donoghue, B.~R.~Holstein and D.~Wyler,
  %``Electromagnetic selfenergies of pseudoscalar mesons and Dashen's theorem,''
  Phys.\ Rev.\  D {\bf 47}, 2089 (1993).
  %%CITATION = PHRVA,D47,2089;%%

\end{thebibliography}
\end{document}